\documentclass[aps,prd,twocolumn,groupedaddress]{revtex4}
\usepackage[dvips]{graphicx}
\usepackage{color}
\usepackage{amsmath,amssymb}
\usepackage{dcolumn}
\usepackage{array}

\begin{document}
\title{Acoustic causality in relativistic shells}

\author{Shunichiro Kinoshita, Yuuiti Sendouda, and Keitaro Takahashi}

\affiliation{
Department of Physics, University of Tokyo, 
7-3-1 Hongo, Bunkyo, Tokyo 113-0033, Japan
}

\begin{abstract}
The motion of sound waves propagating in the perfect fluid with
inhomogeneous background flow is effectively described 
as a massless scalar field on a curved space-time.
This effective geometry is characterized by the acoustic metric, which
depends on the background flow, and null geodesics on the geometry
 express the acoustic causal structure. Therefore by the effective
 geometry we can easily study the causality on the flows.
In this paper, we consider a spherically symmetric, relativistic outflow
and present the maximal causally connected region for a super-sonic flow.
When Lorentz factor of the radial velocity of the flow is constant or
obeys power-law with respect to the radial coordinate $r$, we can solve
 it analytically.
As a result we show that in the constant case the maximum angle is
 proportional to inverse of Lorentz factor and logarithmically increases
 with respect to $r$, in contrast, accelerative expansions in power-law
 case make this angle bounded.
\end{abstract}
\maketitle

\section{Introduction}
Recently a highly polarized prompt $\gamma$-ray emission was reported by
RHESSI observation of the $\gamma$-ray burst (GRB) GRB021206
\cite{Coburn}.
This result has inspired many discussions;
some authors contravene this analysis of the observation \cite{Rutledge}
and others suggested various models to produce such a high polarization
\cite{Waxman,Granot,Lyutikov,Lyutikov2,Nakar}.
For instance, in the electromagnetic models considering Poynting flux
dominated flows, polarization arises from a uniform, large scale
magnetic field \cite{Lyutikov,Lyutikov2}.
On the other hand, in the hydrodynamic models
(e.g., the fireball model \cite{Piran}) it was suggested that
polarization arises even from a random magnetic field \cite{Waxman}.
Besides polarized emissions have been detected also in the GRB afterglows
and there are many discussions whether ordered magnetic fields exist or
not \cite{Covino,Gruzinov,Matsumiya,Sari,Medvedev,Rossi}.

If polarizied emission needs large scale coherence of magnetic fields,
acoustic causality inside the outflow is important to maintain the
coherence.
The acoustic causality of a magnetic-dominated outflow was investigated
in electromagnetic models of GRBs \cite{Lyutikov,Lyutikov2}.
However these analyses were not enough because the effect on the waves
propagating in inhomogeneous flows was neglected.

For studying such propagations in inhomogeneous flows
it is very useful to utilize the effective geometry.
The reason is as follows. Sound wave propagation in an inhomogeneous
flow of a perfect fluid or that of electromagnetic waves in an
inhomogeneous medium are both equivalent to propagating massless scalar
fields on an effective curved space-time.
These effective geometries are determined by background quantities
including the real geometry of spacetime.
This idea has been applied to various systems: fluids (ordinary
\cite{Unruh} or super \cite{Volovik}), dielectrics \cite{Reznik},
non-linear electromagnetism \cite{Novello}, and Bose-Einstein condensates
\cite{Garay,Liberati}.
Furthermore effective geometries permit us to study non-gravitational
systems with methods and ideas of general relativity, such as geodesics,
light cone, black holes, ergo-spheres, Hawking radiation, and so on
\cite{Unruh,Visser,Volovik,Garay,Liberati,Schutzhold,Leonhardt,Sakagami,Barcelo,Novello,Reznik}.
Particular applications in astrophysics are non-linearmagnetism near
a magneter and hydrodynamic accretion-flow onto a black
hole \cite{Bergliaffa,Moncrief}. 

In this paper, we apply the effective geometry to the causality analysis
of relativistic outflows from high-energy sources, which occurs,
for example, with GRBs.
In Sec.~\ref{sec:formalism} we review the acoustic metric associated
with the effective geometry for sound propagation in a relativistic
fluid \cite{Moncrief,Bilic}.
In Sec.~\ref{sec:analysis}, by the effective geometry we shall
analyse the causal structure of a spherical outflow in case that radial
Lorentz factor is constant or power-law with respect to the radial
coordinate.

\section{Acoustic metric\label{sec:formalism}}

We consider the sound wave equation for a perfect, irrotational fluid.
We denote by $u_\mu$, $\rho$, $p$, and $n$ the 4-velocity, energy
density, pressure, and number density of the fluid, respectively.
The energy-momentum tensor of the perfect fluid is given by
\begin{equation}
T_{\mu\nu} = (\rho + p) u_\mu u_\nu + pg_{\mu\nu},\quad u_\mu u^\mu = -1,
\end{equation}
where $g_{\mu\nu}$ is the physical space-time metric with the signature
$(-++ +)$.
The basic equations of the fluid dynamics are the continuity equation
\begin{equation}
\nabla_\mu (nu^\mu) = 0,\label{continuity}
\end{equation}
and the energy conservation
\begin{equation}
\nabla_\mu T^{\mu\nu} = 0.\label{conservation}
\end{equation}
If we assume that the flow is isentropic and irrotational, we can
introduce a scalar function $\varphi$ such that
\begin{equation}
u_\mu = \frac{1}{h}\nabla_\mu \varphi,
\end{equation}
where the specific enthalpy $h$ is defined as
\begin{equation}
h = \frac{\rho+p}{n}.
\end{equation}
In this case, Eqs.~(\ref{continuity}) and (\ref{conservation}) are
rewritten as the two equations for $\varphi$ \cite{Moncrief,Bilic}:
\begin{equation}
\nabla_\mu (\frac{n}{h}g^{\mu\nu}\nabla_\nu \varphi) = 0,
\end{equation}
\begin{equation}
h^2 = -g^{\mu\nu}\nabla_\mu \varphi\nabla_\nu \varphi.
\end{equation}

Next we linearize these equations for small perturbations around the
background flow.
As a result we obtain the wave equation
\begin{equation}
\nabla_\mu \left\{\frac{n}{h}\left[g^{\mu\nu}-(1-\beta_\mathrm
			      s{}^{-2})u^\mu
			      u^\nu\right]\right\}\nabla_\nu \psi =
0,\label{wave_eq}
\end{equation}
where $\psi$ is the perturbations of $\varphi$ and $\beta_\mathrm s$ is
the sound speed given by
\begin{equation}
\beta_\mathrm s{}^2 = \frac{\partial p}{\partial \rho} =
 \frac{n}{h}\frac{\partial h}{\partial n}.
\end{equation} 

Here we define the acoustic metric tensor $G_{\mu\nu}$ and its inverse
$G^{\mu\nu}$ such that
\begin{equation}
G_{\mu \nu} = \frac{n}{h \beta_\mathrm s}\left[(1 - \beta_\mathrm
					  s{}^2)u_\mu u_\nu + g_{\mu
					  \nu} \right], \label{acoustic}
\end{equation}
\begin{equation}
G^{\mu \nu} = \frac{h \beta_\mathrm s}{n}\left[g^{\mu \nu} -
					  (\beta_\mathrm
					  s{}^{-2}-1)u^\mu u^\nu
					 \right].
\end{equation}
Then the wave equation (\ref{wave_eq}) is rewritten as
\begin{equation}
\frac{1}{\sqrt{G}}\partial_\mu (\sqrt{G}G^{\mu \nu} \partial_\nu \psi) =
 0,
\end{equation}
where
\begin{equation}
\sqrt{G} \equiv \sqrt{-\det G_{\mu\nu}} = \frac{n^2}{h^2 \beta_\mathrm
 s}\sqrt{-\det g_{\mu\nu}}.
\end{equation}
This is the d'Alembert equation on the curved space-time with the metric
$G_{\mu\nu}$.
Thus the sound wave propagating in a perfect fluid is
described as a massless scalar field on the effective geometry
determined by the background flow. 

\section{Causality analysis\label{sec:analysis}}
\subsection{Spherically symmetric outflow}

We consider a spherically symmetric, stationary outflow.
The fluid 4-velocity, $u_\mu$, is then
\begin{equation}
u_\mu = \gamma (-\mathrm dt + \beta \mathrm dr),
\end{equation}
where $\beta$ and $\gamma$ are its radial velocity and Lorentz factor,
respectively.
Because of the stationarity, $\beta$ depends only on the radial
coordinate $r$.
We assume that the gravity is negligible and the physical space-time is
flat. 
The acoustic metric is
\begin{align}
G_{\mu\nu} &\propto - \left(1-\frac{\gamma^2}{\gamma_\mathrm s
 {}^2}\right)\mathrm dt^2 - 2 \frac{\gamma^2}{\gamma_\mathrm s {}^2}
 \beta \mathrm dt \mathrm dr \nonumber
\\&\qquad + \left(\beta_\mathrm s {}^2+\frac{\gamma^2}{\gamma_\mathrm s
 {}^2}\right)\mathrm dr^2 + r^2\mathrm d\Omega^2\\
&\propto -(1-\alpha^2)\mathrm dt_*{}^2 + \frac{\beta_\mathrm s
 {}^2}{1-\alpha^2} \mathrm dr^2 + r^2\mathrm d\Omega^2,
\end{align}
where
\begin{equation}
\alpha \equiv \frac{\gamma}{\gamma_\mathrm s},\quad \mathrm dt_* \equiv
 \mathrm dt + \frac{\alpha^2}{1-\alpha^2}\beta \mathrm dr ,\label{def_t}
\end{equation}
and $\gamma_\mathrm s$ is the Lorentz factor of the sound speed.
It is worth mentioning that $\alpha$ is a parameter showing whether the
flow is super-sonic or sub-sonic. 

Since sound waves propagate as massless scalar fields on this metric, 
we consider null geodesics in order to discuss the acoustic causality.
The Lagrangian is
\begin{equation}
2L = -(1-\alpha^2)\dot{t}_*{}^2 + \frac{\beta_\mathrm s {}^2}{1-\alpha^2}
 \dot{r}^2 + r^2\dot{\theta}^2 + r^2 \sin\theta \dot{\phi}^2,
\end{equation} 
where $\cdot\equiv \mathrm d/\mathrm d\lambda$ and $\lambda$ is an
affine parameter.
We neglect the conformal factor of $G_{\mu\nu}$ because it is not
important for null geodesics \cite{Wald}.
Because of the spherical symmetry, we shall consider only the $\phi =
\text{const.}$ case.
By introducing $l$ as the constant of motion, null geodesic equations
become (a) for super-sonic case ($\alpha > 1$) :
\begin{align}
\frac{\mathrm d\theta}{\mathrm dt_*} &= \frac{l
 \sqrt{\alpha^2-1}}{r^2}\label{super1},\\
\frac{\mathrm dr}{\mathrm dt_*} &= \pm\frac{\alpha^2 -1}{\beta_\mathrm
 s}\sqrt{1+\frac{l^2}{r^2}}\label{super2}\qquad(-\infty < l < \infty),
\end{align}
where $l$ corresponds with ``angular momentum'' related to an initial
emitting angle of a null geodesic,
and (b) for sub-sonic case ($\alpha < 1$) :
\begin{align}
\frac{\mathrm d\theta}{\mathrm dt_*} &= \frac{l
 \sqrt{1-\alpha^2}}{r^2}\label{sub1},\\
\frac{\mathrm dr}{\mathrm dt_*} &= \pm\frac{1-\alpha^2}{\beta_\mathrm
 s}\sqrt{1-\frac{l^2}{r^2}}\label{sub2}\qquad(-r_0 < l < r_0),
\end{align}
where $r_0$ is the radius of a point from which the sound wave is
emitted, in this case $l$ is bounded.

The hypersurface generated by these null geodesics for all $l$ is the 
``light cone'' of the sound.
Therefore the inside of this surface is the causally connected region
with the sound wave.

Now we shall analyse the causality in two simple examples of outflows:
the Lorentz factor of the radial velocity is (i) constant
and (ii) power-law with respect to the radius.
These situations often appear in various models of GRB
\cite{Piran,Lyutikov2,Lyutikov,Dar}.
Even in more general cases, the analysis will be similar to what is
shown below.

\subsection{Relativistic fluid with a constant velocity}
Here, for simplicity, we assume that the radial velocity of fluid,
$\beta$, is a constant and that the fluid equation of state is
relativistic and independent of $r$, i.e., the sound speed,
$\beta_\mathrm s = 1/\sqrt{3}$, is also a constant.
Now we firstly assume that a shell is infinitely thick
so that it can be dealt as a steady outflow.

In the $\alpha>1$ case (super-sonic) the solutions of
Eqs.~(\ref{super1}) and (\ref{super2}) are
\begin{equation}
r = \sqrt{\left(\frac{\alpha^2-1}{\beta_\mathrm s}\right)^2t_*{}^2
\pm 2\frac{\alpha^2-1}{\beta_\mathrm
s}\sqrt{l^2+r_0{}^2}t_*+r_0{}^2}\label{r},
\end{equation}
and
\begin{equation}
\theta = \frac{\beta_\mathrm s}{\sqrt{\alpha^2-1}}\ln
\frac{r(l+\sqrt{l^2+r_0{}^2})}{r_0(l+\sqrt{l^2+r^2})}\label{theta},
\end{equation}
where $t_* = t - \frac{\alpha^2}{\alpha^2-1}\beta (r-r_0)$.
Here the positive sign represents outgoing geodesics, while the negative
ingoing geodesics. 
Eliminating $l$ from these two equations, we obtain the equation for $t$,
$r$, and $\theta$, which describes the ``light cone'' of sound.
For a given $t$, the cross section of that surface becomes the boundary
of causally connected region. 

\begin{figure}[h]
\begin{center}
\includegraphics[width=8cm]{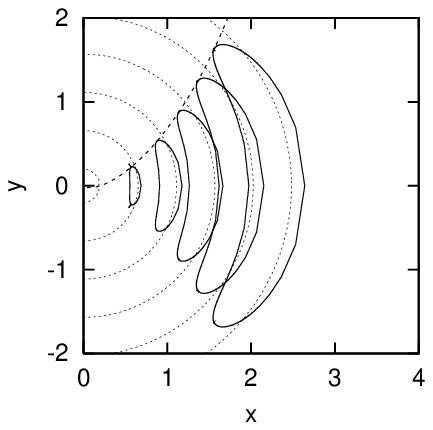}
\caption{The propagation of the sound wave for the super-sonic case with
 $\alpha=2$ (solid curve).
For $\alpha=2$ the velocity of the fluid is mildly relativistic
 ($\gamma=\sqrt{6}$).
The dashed curve is the maximally spread trajectory in the
 causally connected region.
The dotted curves express the evolution of shell on which the sound wave
 initially emitted.
The initial radius of the shell is taken as $r_0=0.2$.}\label{fig:2}
\includegraphics[width=8cm]{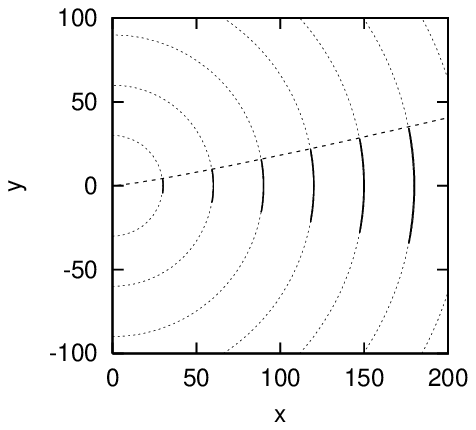}
\caption{The propagation of the sound wave for the super-sonic case with
 $\alpha=20$ (solid curve).
The velocity of the fluid is ultra-relativistic with $\gamma=10\sqrt{6}$.
The dashed curve is the maximally spread trajectory in the causally
 connected region.
The dotted curves express the evolution of shell on which the sound wave
 initially emitted.
In comparison with FIG.~\ref{fig:2} the sound wave does not spread
 radially and propagates almost on the shell.
The initial radius of the shell is taken as $r_0=0.2$.}\label{fig:20}
\includegraphics[width=8cm]{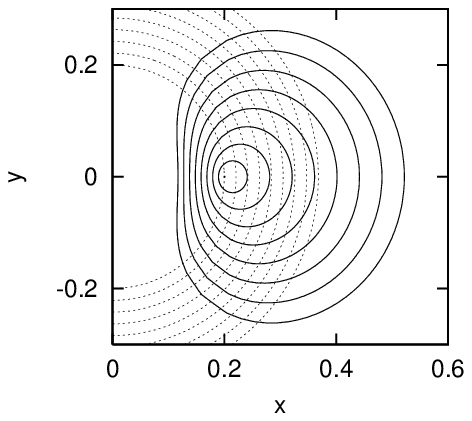}
\caption{The propagation of the sound wave for the sub-sonic case with
 $\alpha=0.9$ (solid curve).
The dotted curves express the evolution of shell on which the sound wave
 initially emitted.
The initial radius of the shell is taken as $r_0=0.2$.}\label{fig:09}
\end{center}
\end{figure}

The geodesic with $l\to\infty$ realizes the maximum value of polar angle.
In the limit $l \gg r$, the trajectory is 
\begin{equation}
\theta_\mathrm{max}(r) \simeq \frac{\beta_\mathrm s}{\sqrt{\alpha^2 -1}}
 \ln \frac{r}{r_0}\label{max_angle}.
\end{equation}
For a given $r$ the region with polar angle larger than
$\theta_\mathrm{max}(r)$ is causally disconnected from a source emitting
the sound wave.
The propagation of the sound wave and the trajectory with the maximum
angle in a mildly relativistic case ($\alpha=2$)
and an ultra-relativistic case ($\alpha=20$) are shown in
FIG.~\ref{fig:2} and FIG.~\ref{fig:20}, respectively.
We found that the maximally causal polar angle is $\sim \gamma^{-1}$,
and for an outflow with a constant velocity it becomes unboundedly large 
as $r$ extends (i.e., as time increases).

Similarly, in the $\alpha<1$ case (sub-sonic) we can see the causally
connected region by solving the null geodesic equations (\ref{sub1}) and
(\ref{sub2}).
The propagation of sound wave with $\alpha=0.9$ is given in
FIG.~\ref{fig:09}.
The solutions are slightly different from those of the super-sonic case
 in the sign of $\alpha^2-1$.

Another example is the thin-shell approximation.
If we assume that a sound wave can propagate only on an infinitely
thin shell, it satisfies a condition $\mathrm dr=\beta \mathrm dt$.
Thus the geodesic equations are now
\begin{align}
\frac{\mathrm d\theta}{\mathrm dt} &= \pm \frac{\beta_\mathrm
 s}{\gamma}\frac{1}{r},\\
\frac{\mathrm dr}{\mathrm dt} &= \beta.
\end{align}
Therefore the maximum polar angle that the sound wave emitted at $r=r_0$
can reach is 
\begin{equation}
\theta_\mathrm{max}(t) = \frac{\beta_\mathrm s}{\gamma\beta}\ln
 \frac{r(t)}{r_0},
\end{equation}
where $r(t)=r_0+\beta t$.
This result is quite similar to Eq.~(\ref{max_angle}).


\subsection{Power-law case}
In this section, let us consider that the radial Lorentz factor obeys
power-law, such as $\gamma \propto r^p$.
For $p>0$ the expansion velocity of the flow is accelerated and for $p<0$
decelerated.
Now we are interested in the maximum angle in the ultra-relativistic case
so we shall restrict our discussion to the case $\alpha\gg 1$ and $l\gg r$.
Eqs.~(\ref{super1}) and (\ref{super2}) read
\begin{equation}
\frac{\mathrm d\theta}{\mathrm dr} \simeq \frac{\beta_\mathrm s}{\alpha
 r} \propto r^{-p-1},
\end{equation}
and integrating it results in
\begin{equation}
\theta_\mathrm{max} \propto r^{-p} + \text{const}.
\end{equation}
This implies that if
$p>0$, this angle will approach to a constant value
at late time, i.e., the region corresponding polar angles larger than
this constant value will never be causally connected.
In terms of the effective geometry, this acoustic metric has an event
horizon similar to that of de Sitter space in the cosmology \cite{H_E}.
On the other hand, if $p<0$ the maximum angle spreads continuously.

\section{Summary and discussion\label{sec_summary}}
We have studied the acoustic causality in relativistic spherical
outflows with constant and power-law Lorentz factor case by means of the
effective geometry.
The effective metric made the analysis of a sound propagation on
inhomogeneous background flows very easy.

When the Lorentz factor of the outflow is constant,
the maximum polar angle of the causally connected region is proportional
to the inverse of the Lorentz factor of the outflow and
the logarithm of the radius.
For a constant or decelerative expansion the maximum angle increases as
time without a bound. However for accelerated expansion this angle will
reach a constant value at late time and this means the region outside
this angle is never causally connected. These results are very similar to
the expanding universe in the cosmology.

If a shell has a finite thickness sound waves may reach the edge of the
shell.
For a given thickness we can consider only the region between both edges
of the shell.
Therefore the previous trajectory with the maximal polar angle is
outside the shell.
In this case the causally connected region will be narrower than the
case without edges.

In this paper, we dealt with simple examples.
However if a model of GRBs is given and background flow is determined
we can investigate causality in a similar way.
We expect that even in MHD or fluids with vorticity these results are
basically unchanged \cite{Bergliaffa2}.

\section*{Acknowledgement}
We thank Masaki Sano, Hiroyuki Yoshiguchi, Tomoya Takiwaki, and Kazuhiro
Yahata for helpful discussions.
The work of YS and KT is supported by a Grant-in-Aid from JSPS.

\bibliography{0405149}

\end{document}